\documentclass[twoside,twocolumn,reprint,secnumarabic,superscriptaddress,amssymb,amsmath,nobibnotes,nofootinbib,aps,floatfix]{revtex4-1}
\usepackage{mathptmx}

\usepackage[latin9]{inputenc}
\usepackage{color}
\usepackage{amsmath}
\usepackage{graphicx}
\usepackage[unicode=true,pdfusetitle,
 bookmarks=true,bookmarksnumbered=false,bookmarksopen=false,
 breaklinks=true,pdfborder={0 0 0},pdfborderstyle={},backref=false,colorlinks=true]
 {hyperref}
\hypersetup{
 linkcolor=blue,citecolor=blue,urlcolor=blue}

\makeatletter
\graphicspath{{figures/}}
\usepackage{float}
\usepackage{tikz}
\usetikzlibrary{calc,shapes,backgrounds}
\usetikzlibrary{datavisualization,patterns}
\usetikzlibrary{arrows.meta}
\usepackage{pslatex}
\usepackage{color}

\usepackage{array}
\newcommand{\PreserveBackslash}[1]{\let\temp=\\#1\let\\=\temp}
\newcolumntype{C}[1]{>{\PreserveBackslash\centering}p{#1}}
\newcolumntype{R}[1]{>{\PreserveBackslash\raggedleft}p{#1}}
\newcolumntype{L}[1]{>{\PreserveBackslash\raggedright}p{#1}}

\makeatother

\begin{document}
\title{Critical properties of the Susceptible-Exposed-Infected model with
correlated temporal disorder}
\author{Alexander H. O. Wada}
\affiliation{Instituto de Física de São Carlos, Universidade de São Paulo, C.\ P.\ 369,
São Carlos, São Paulo 13560-970, Brazil.}
\author{José A. Hoyos}
\affiliation{Instituto de Física de São Carlos, Universidade de São Paulo, C.\ P.\ 369,
São Carlos, São Paulo 13560-970, Brazil.}
\begin{abstract}
In this paper we study the critical properties of the non-equilibrium
phase transition of the Susceptible-Exposed-Infected model under the
effects of long-range correlated time-varying environmental noise
on the Bethe lattice. We show that temporal noise is perturbatively
relevant changing the universality class from the (mean-field) dynamical
percolation to the exotic infinite-noise universality class of the
contact process model. Our analytical results are based on a mapping
to the one-dimensional fractional Brownian motion with an absorbing
wall and is confirmed by Monte Carlo simulations. Unlike the contact
process, our theory also predicts that it is quite difficult to observe
the associated active temporal Griffiths phase in the long-time limit.
Finally, we also show an equivalence between the infinite-noise and
the compact directed percolation universality classes by relating
the SEI model in the presence of temporal disorder to the Domany-Kinzel
cellular automaton in the limit of compact clusters.
\end{abstract}
\date{\today}
\maketitle

\section{Introduction}

Absorbing phase transitions are non-equilibrium phase transitions
separating an inactive phase without fluctuations (absorbing) from
an active (fluctuating) one. This type of phase transition is observed
in many models for a great variety of phenomena such as epidemic spreading~\citep{Harris_CP,PGrassberger_GEP-DyP},
interface growth~\citep{ALBarabasi_Surface}, catalytic chemical
reactions~\citep{ZGB_model}, turbulent crystal liquids~\citep{Chate_PRL2007,Chate_DPPRE},
periodically driven suspensions~\citep{LCorte_ROPDS2008,Franceschini_PRL2011},
superconducting vortices~\citep{SOkuma_TfromRtoIR2011} and bacteria
colony biofilms~\citep{KSKorolev_CC2011,KSKorolev_PG2011}. (See,
e.g., Refs.\ \citealp{MarroDickman99,Hinrichsen00,Odor04,HenkelHinrichsenLuebeck_book08}
for reviews). Recently, absorbing state phase transitions have also
been observed in open driven many-body quantum systems~\citep{Gutierrez_AbsPhaseTransManyBodyQuantum}.

An important and thoroughly studied universality class of absorbing
phase transitions is the ``ubiquitous'' directed percolation (DP)
universality class~\citep{grassberger-delatorre-ap79,janssen-zpb81,grassberger-zpb82,MarroDickman99,Hinrichsen00,Odor04,HenkelHinrichsenLuebeck_book08}.
For many years it has evaded experimental confirmation~\citep{Chate_PRL2007,NatPhys_DPinCouetteflow,NatPhys_DPsimul+exp,NatPhys_DPinGel}.
Earlier, it was thought that one possible reason was due to quenched
disorder as the DP universality class is perturbatively unstable against
it for $d<4$~\citep{Harris_Criterion,kinzel-zpb85,Noest86}. It
was then determined that quenched disorder in the contact process~\citep{Harris_CP}
(a prototypical model exhibiting a transition in the DP universality
class) induces a rich physical scenario with a universal (disorder-independent)
infinite-randomness critical point governing the critical properties,
in addition to off-critical Griffiths singularities in the inactive
phase~\citep{HooyberghsIgloiVanderzande03,Hooyberghs_DCPPRE2004,VojtaDickison05,Hoyos08,SCF_2DCPQuenched,VojtaFarquharMast09,Vojta12,Vojta_HCRareRegions,Vojta_5dDCP,AHOW_DCP}.
For a review on the exotic properties of the infinite-randomness universality
class and the corresponding Griffiths phase, see, e.g., Ref.~\citealp{Vojta06}.
Despite of this theoretical achievement, an experimental confirmation
of this exotic critical behavior is still lacking in non-equilibrium
phase transitions.

The effects of time-dependent global fluctuations (temporal disorder)
has also attracted attention~\citep{Leigh81,kinzel-zpb85,jensen-prl96,jensen-jpa-05,kamenev-etal-prl08,OvaskainenMeerson10,TemporalGP_Vaszquez,VojtaHoyos15,BarghathiVojtaHoyos16,Fiore_SecondOrderTempDis2016,Fiore_SecondOrderTempDis2018}.
Like its spatial counterpart, it is a relevant perturbation for the
DP critical behavior~\citep{kinzel-zpb85}, except that it is relevant
in all dimensions. The physical behavior is equally rich as well.
An infinite-noise critical point governs the transition in all dimensions~\citep{VojtaHoyos15,BarghathiVojtaHoyos16}
(implying ever increasing relative density fluctuations) and an associated
active Griffiths phase also exists~\citep{TemporalGP_Vaszquez}.
In $d=\infty,$ the infinite-noise critical point is akin to the infinite-randomness
critical point with the roles of time and space exchanged. It turns
out that the relative density fluctuations increase unbounded $\sim t^{1/2}$.
For finite $d$, the infinite-noise critical point is qualitatively
different, and is akin to the disordered Kosterlitz-Thouless critical
point in dissipative quantum rotors~\citep{vojta-etal-jp11} and
in the quantum Ising model with long-range couplings~\citep{juhasz-kovacs-igloi-epl14}.
In addition, the relative density fluctuations increase only $\sim\ln t$.
Later on, the effects of correlated noise was also investigated in
the $d=\infty$ limit~\citep{AHOW_ExtinctionTransition}. It was
observed an increase (decrease) of the critical and off-critical fluctuations
when the correlations are positive (negative). 

Given this ``ubiquitous'' infinite-noise criticality, it is desirable
to know whether a similar scenario also happens in other non-equilibrium
phase transition which are not in the DP universality class. We then
study the Susceptible-Exposed-Infected (SEI) model which exhibits
a non-equilibrium phase transition into an absorbing state in the
dynamical percolation universality class~\citep{SEI_2011,SEI_2015}.
Considering the model on the Bethe lattice, we could solve the effects
of correlated temporal disorder by mapping the problem onto a fractional
Brownian motion with an absorbing wall (fBMAW). We show that this
model exhibits the same infinite-noise critical behavior as in the
($d=\infty$) contact process, but very distinct off-critical Griffiths
phase. In addition, our analytical results are confirmed by Monte
Carlo simulations. Finally, we also uncover an equivalence between
the infinite-noise and the compact directed percolation universality
classes. This is done by a one-to-one correspondence between the critical
dynamics of the SEI model and of the Domany-Kinzel cellular automaton~\citep{CDP_DomanyKinzel1984}
in the limit of compact clusters . Therefore, we show that the notion
of infinite-noise criticality is more common than previously thought.

This paper is organized as follows. We introduce the SEI model with
temporal disorder and the mapping onto fBMAW in Sec.\ \ref{SEC:SEI}.
In Sec.\ \ref{SEC:Theory} we develop our theory and derive the critical
behavior of key observable in order to characterize the universality
class. The Monte Carlo simulations confirming our theory are presented
in Sec.\ \ref{SEC:MC}. Finally, we uncover the equivalence between
the compact directed percolation and the infinite-noise universality
classes in Sec.~\ref{sec:CDP}, and conclude in Sec.\ \ref{SEC:Conclusion}
with final remarks.

\section{Susceptible-Exposed-Infected model \label{SEC:SEI}}

\subsection{Definition \label{SEC:SEI-def}}

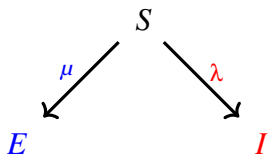
\begin{figure}[b]
\centering

\begin{tikzpicture}
	\node at (2.03,2) {\large \color{black}\emph{S}} ;
	\draw[very thick, ->] (1.7, 1.7) -- (0.7, 0.7);
	\node[left] at (1.2, 1.3) { \color{blue}\textbf{$\mu$}};
	\draw[very thick, ->] (2.3, 1.7) -- (3.3, 0.7);
	\node[right] at (2.8, 1.3) { \color{red}\textbf{$\lambda$}};
	\node at (0.35, 0.35) {\large \color{blue}\emph{E}};
	\node at (3.6, 0.35) {\large \color{red}\emph{I}};
\end{tikzpicture} \caption{Reactions of the susceptible-exposed-infected (SEI) model. In the
presence of an infected nearest neighbor, susceptible individuals
becomes either exposed or infected with rates $\mu$ and $\lambda$,
respectively.\label{FIG:SEI}}
\end{figure}

The Susceptible-Exposed-Infected (SEI) model~\citep{SEI_2011} was
introduced to describe the spreading of a certain disease on a susceptible
population. In this model individuals are lattice sites and the interactions
occur only between infected (\emph{I}) and susceptible (\emph{S})
nearest neighbor individuals. When an interaction takes place, the
\emph{S} individual either becomes exposed (\emph{E}) or infected
\emph{(I)} with rates $\mu$ and $\lambda$ respectively. We emphasize
that neither \emph{E} or \emph{I} individuals are allowed to change
their state, and \emph{E} individuals do not spread the disease. Figure
\ref{FIG:SEI} shows the possible reactions.

Therefore, the system is active whenever the number $n_{IS}$ of susceptible-infected
pairs (active bonds) is nonzero. In a finite regular lattice, the
SEI model will eventually reach an absorbing state characterized by
a cluster of \emph{I} sites surrounded by $E$ ones. Notice that the
model admits an infinite number of absorbing states.

We now explain the simulation algorithm according to the Gillespie
method~\citep{Gillespie}. We maintain a list of the $n_{IS}$ active
bonds and proceed as follows: (i) increase time by $1/(n_{IS}r)$,
where $r$ is a uniformly distributed random number in the interval
$]0,1[$; (ii) select, at random and with equal probability, one active
bond in the list; (iii) the \emph{S} site of the selected bond either
becomes \emph{E} with probability $q=\mu/(\mu+\lambda)$, or becomes
\emph{I} with probability $p=1-q$; (iv) update the list of active
bonds; (v) return to (i) until there are no active bonds. 

Interestingly, the above algorithm reveals a direct connection between
the SEI model and the isotropic site percolation problem. Consider,
for instance, a lattice in which all sites are initially in state
\emph{S} except for one which is in the \emph{I} state. At the end
of the simulation, the Gillespie algorithm will have generated a cluster
of \emph{I} sites surrounded by \emph{E} ones where each of them were
asked just once whether they are at the \emph{E} (vacant) or \emph{I}
(occupied) state, just like in the isotropic percolation problem~\citep{PercolationBook}.
(The remaining \emph{S} sites which did not participate in the simulation
are not important here.) Therefore, the SEI model exhibits a phase
transition in the dynamical percolation universality class. For $p<p_{c}$
(the isotropic percolation threshold), the generated \emph{I} cluster
is finite (inactive phase), whereas an infinite one is possible for
$p>p_{c}$ (active phase). Furthermore, the static critical exponents
such as $\beta$ and $\nu_{\perp}$ are the same as those of isotropic
percolation.\footnote{See Ref.\ \citealp{SEI_2015} for the critical properties of the
SEI model on a square lattice, and Ref.\ \citealp{AHOW_DCP} for
its application to percolation.}

\subsection{Solution on the Bethe lattice}

We now review the solution of the SEI model on a infinite Bethe lattice~\citep{SEI_2011}
restricting ourselves to the initial condition of a single cluster
of \emph{I} sites in a lattice filled of \emph{S} sites. In this case,
any arbitrary cluster configuration containing $n_{I}$ infected sites
will have a number of individuals on its perimeter that only depends
on $n_{I}$. In other words, although it is possible to rearrange
the connected cluster of \emph{I} individuals in many different ways,
the number of sites on the perimeter depends only on $n_{I}$ regardless
of the particular cluster rearrangement. In fact, the number exposed
individuals $n_{E}$, $n_{IS}$ and $n_{I}$ are related via 
\begin{equation}
n_{IS}+n_{E}=(z-2)n_{I}+2,\label{EQ:perimeter-bulk}
\end{equation}
where $z$ is the coordination number. Notice that the left hand side
is the number of sites at the perimeter of the cluster of \emph{I}
individuals. 

The time evolution of the average number of infected, exposed and
active bonds, $N_{I}=\langle n_{I}\rangle$, $N_{E}=\langle n_{E}\rangle$
and $N_{IS}=\langle n_{IS}\rangle$, where $\langle\ldots\rangle$
is the average over the stochastic noise, can be obtained from the
following differential equations: 
\begin{align}
\frac{\text{d}}{\text{d}t}N_{I} & =\lambda N_{IS},\\
\frac{\text{d}}{\text{d}t}N_{E} & =\mu N_{IS},\\
\frac{\text{d}}{\text{d}t}N_{IS} & =-\left(\mu+\lambda\right)N_{IS}+(z-1)\lambda N_{IS},\label{eq:NIS-dif}
\end{align}
where we have redefined $\lambda\rightarrow z\lambda$ and $\mu\rightarrow\mu z$
for convenience. These equations can be solved exactly yielding to
\begin{align}
N_{I} & =\frac{\mu_{c}N_{IS}^{0}}{(z-2)\left(\mu_{c}-\mu\right)}\left(e^{(\mu_{c}-\mu)t}-1\right)+N_{I}^{0},\label{EQ:NI-clean}\\
N_{E} & =\frac{\mu N_{IS}^{0}}{\mu_{c}-\mu}\left(e^{(\mu_{c}-\mu)t}-1\right)+N_{E}^{0},\label{EQ:NE-clean}\\
N_{IS} & =N_{IS}^{0}e^{(\mu_{c}-\mu)t},\label{EQ:NIS-clean}
\end{align}
 with 
\begin{equation}
\mu_{c}=\left(z-2\right)\lambda.\label{eq:muc}
\end{equation}
Note that Eq.~\eqref{EQ:perimeter-bulk} is satisfied provided that
$N_{IS}^{0}+N_{E}^{0}=(z-2)N_{I}^{0}+2$. 

There are three distinct behaviors. The active phase ($\mu<\mu_{c}$)
is characterized by a diverging $N_{IS}$ as $t\rightarrow\infty$
whereas in the inactive phase ($\mu<\mu_{c}$) $N_{IS}\rightarrow0$
as $t\rightarrow\infty$. The critical point happens for $\mu=\mu_{c}$
where $N_{IS}$ is constant in time, and $N_{I}$ and $N_{E}$ grow
only linearly with $t$.\footnote{The probability of an infected site $p=\lambda/(\mu+\lambda)$ equals
to $p_{c}=1/(z-1)$ at criticality, which is the percolation threshold
for the Bethe lattice~\citep{PercolationBook}. }

\subsection{Temporal disorder and the mapping onto a fractional Brownian motion
with an absorbing wall \label{SEC:TD-ABS}}

We are interested in the situation where the system is under the influence
of environmental noise. Following Ref.\ \citealp{VojtaHoyos15},
we introduce temporal disorder by allowing the rates $\mu$ and $\lambda$
to change randomly after remaining constant during a time interval
of size $\Delta t_{n}>0$. Thus, the rates at the $n$th time interval
are $\mu_{n}=\mu+\Delta\mu_{n}$ and $\lambda_{n}=\lambda+\Delta\lambda_{n}$
with $\lambda$ and $\mu$ being constants and $\Delta\mu_{n}$ and
$\Delta\lambda_{n}$ being zero-average noises.

Given the sequence $\left\{ \lambda_{n},\mu_{n}\right\} $, the complete
time evolution of $N_{IS},$ $N_{E}$ and $N_{I}$ is thus obtained
by patching the solutions \eqref{EQ:NI-clean}, \eqref{EQ:NE-clean}
and \eqref{EQ:NIS-clean} in each time interval. The solution for
$N_{IS}$ reads 
\begin{align}
N_{IS}^{(n+1)} & =N_{IS}^{(n)}e^{\left(\mu_{c,n}-\mu_{n}\right)\Delta t_{n}},\label{EQ:NIS-dirty}
\end{align}
where $N_{IS}^{(n)}$ is the stochastic average of $n_{IS}$ at the
beginning of the $n$th time interval and $\mu_{c,n}\equiv(z-2)\lambda_{n}$.
It is convenient to recast Eq.\ \eqref{EQ:NIS-dirty} in terms of
$x_{n}=\ln(1+N_{IS}^{(n)})$: $x_{n+1}=\ln\left(1+\left(e^{x_{n}}-1\right)e^{\left(\mu_{c,n}-\mu_{n}\right)\Delta t_{n}}\right)$.
In the $x_{n}\gg1$ limit, the time evolution of $x$ then becomes
a simple random walk 
\begin{equation}
x_{n+1}=x_{n}+\zeta_{n}
\end{equation}
with an absorbing wall at $x=0$ (ensuring $N_{IS}\geq0$) and noise
\begin{equation}
\zeta_{n}=\left(\mu_{c,n}-\mu_{n}\right)\Delta t_{n}=\left((z-2)\lambda_{n}-\mu_{n}\right)\Delta t_{n}.\label{eq:bare-distanceFC}
\end{equation}
The mean bias is thus 
\begin{equation}
v=\left[\frac{\zeta_{n}}{\Delta t_{n}}\right]=\mu_{c}-\mu,\label{EQ:bias-v}
\end{equation}
with $\mu_{c}=\left[\mu_{c,n}\right]=\left(z-2\right)\lambda$ as
in the clean case Eq.~\eqref{eq:muc}. Here, $\left[\cdots\right]$
denotes the average over the temporal disorder. Clearly, $v>0$ means
that $N_{IS}$ typically increases in time and thus the system is
in the active phase. On the other hand, the inactive phase happens
for $v<0$ since the system will eventually reach the absorbing state.
Finally, the transition takes place for $v=0$. 

In order to obtain a complete description of the critical behavior,
we will need the noise correlation function 
\begin{equation}
G(i,j)=[\zeta_{i}\zeta_{j}]-[\zeta_{i}][\zeta_{j}].\label{eq:G}
\end{equation}
 In this work, we will assume a fractional Gaussian noise~\citep{Qian2003}
\begin{equation}
G(i,j)=G(\tau)=\frac{\sigma^{2}}{2}(|\tau+1|^{2-\gamma}-2|\tau|^{2-\gamma}+|\tau-1|^{2-\gamma}),\label{EQ:G-FBM}
\end{equation}
which decays $\sim(1-\gamma)\tau^{-\gamma}$ for large $\tau=|i-j|$.
Here, the correlation exponent $\gamma$ takes any value in the interval
$]0,2[$ and is related to the Hurst exponent $H$ through $2H=2-\gamma$.
This implies that the fractional Gaussian noise is positively correlated
for $\gamma<1$, negatively correlated for $\gamma>1$, and uncorrelated
for $\gamma=1$. Finally, $\sigma^{2}$ is the variance of the noise
$\zeta$.

We end this section by summarizing our results so far. In the large
$N_{IS}$ limit, the SEI model with the correlated temporal disorder
\eqref{EQ:G-FBM} can be mapped onto a fractional Brownian motion
with an absorbing wall (fBMAW) at $x=0$ with each step of the walker
$n\rightarrow n+1$ corresponding to the time increasing $t\rightarrow t+\Delta t_{n}$
in the SEI model. We will make use of this fact to study the system
behavior near the transition. We anticipate that our results are accurate
even though the approximation fails near the absorbing wall. The reason
for such, as in the contact process~\citep{VojtaHoyos15,AHOW_FBM}
and related models~\citep{Fiore_SecondOrderTempDis2018}, is that
details of the wall are irrelevant for the singular critical behavior.

\section{Theory \label{SEC:Theory}}

\subsection{Generalized Harris criterion\label{subsec:Harris}}

Before developing the theory, let us first discuss whether the clean
critical behavior is perturbatively unstable against correlated temporal
disorder. Following Harris~\citep{Harris_Criterion}, we need to
study the distribution $Q(\psi)$ of the coarse-grained distances
from criticality as the critical point is approached. Such probability
distribution is obtained by averaging $\zeta_{n}$ in Eq.~\eqref{eq:bare-distanceFC}
over a time window of order of the clean correlation time $\xi_{t}$.
For $\xi_{t}\rightarrow\infty$, $Q(\psi)$ is known since $\zeta_{n}$
is a fractional Gaussian noise\ \citep{Qian2003}. It is simply a
Gaussian with the mean $\bar{\psi}=[\zeta_{n}]=\zeta$ and variance
$\sigma_{\psi}^{2}=\xi_{t}^{-\gamma}$, where $\gamma$ is the correlation
exponent in Eq.\ \eqref{EQ:G-FBM}. A necessary condition for the
stability of clean critical theory is that $\sigma_{\psi}/\left|\text{\ensuremath{\bar{\psi}}}\right|\rightarrow0$
as $\bar{\psi}=\zeta\rightarrow0$. Since $\zeta\sim\xi_{t}^{-1/\nu_{\parallel}^{(c)}}$,
where $\nu_{\parallel}^{(c)}$ is the correlation time exponent of
the clean theory, then we conclude that the system is unstable against
fractional Gaussian correlated temporal disorder if 
\begin{equation}
\gamma\nu_{\parallel}^{(c)}<2.\label{EQ:GHC}
\end{equation}
As the SEI model belongs to the dynamical percolation universality
class, then $\nu_{\parallel}^{(c)}=1$ for $d\geq6$~\citep{HenkelHinrichsenLuebeck_book08},
therefore disorder is relevant for all $\gamma$ in the Bethe lattice.
For $d<6$ correlated disorder is also a relevant perturbation for
some values of $\gamma$.\footnote{For $d=2,$$\nu_{\parallel}^{(c)}=1.5079(4)$~\citep{SEI_2015}.
Respectively for $d=3,\ 4$ and $5$, it is known that $z=1.375(5),\ 1.605(9)$
and $1.815(10)$~\citep{DyP_CP_HighDim_Hinrichsen2004}, and $\nu_{\perp}^{(c)}=0.8774(13),\ 0.6852(28)$
and $0.5723(18)$~\citep{PercNuPerp_Koza2016}. Since $\nu_{\parallel}^{(c)}=z\nu_{\perp}^{(c)}$,
then $\nu_{\parallel}^{(c)}\approx1.20,\ 1.10$ and $1.04$.}

We end this section by emphasizing that the effects of correlated
disorder can be studied in a completely generic scenario. According
to Ref.~\citealp{VojtaDickman_HarrisCriterion}, the clean critical
behavior is unstable against weak temporal disorder if 

\begin{equation}
\lim_{\xi_{t}\rightarrow\infty}\xi_{t}^{2/\nu_{\parallel}^{(c)}-1}\int_{0}^{\xi_{t}}\text{d}tG(t)\rightarrow\infty,\label{EQ:HC-G}
\end{equation}
where $G(t)$ is the noise correlation function \eqref{eq:G}. Inserting
\eqref{EQ:G-FBM} onto \eqref{EQ:HC-G}, the criterion \eqref{EQ:GHC}
is recovered. However, the integral \eqref{EQ:HC-G} yields two terms
for a more generic power-law correlation such as $G(t)\propto\left(1+\left|t\right|\right)^{-\gamma}$:
one proportional to $\xi_{t}^{1-\gamma}$ and the other is a constant.
(For the fractional Gaussian noise the constant term is zero, i.e.,
the white noise component is vanishing.) This means that a white noise
is present in the latter case but not in the fractional Gaussian noise.
For $\gamma<1$ the first term dominates and the criterion follows
as in Eq.\ \eqref{EQ:GHC}. For $\gamma>1$, on the other hand, the
constant term dominates in the latter case and the criterion changes
to $\nu_{\parallel}^{(c)}<2$. Therefore the system should be unstable
against the generic power-law correlated noise $G\propto(1+\left|t\right|)^{-\gamma}$
if $\text{min}\{1,\gamma\}\nu_{\parallel}^{(c)}<2$.

\subsection{Critical properties}

As discussed in Sec.~\ref{SEC:TD-ABS} when mapping the SEI model
onto a fBMAW, we expect activity to (i) vanish exponentially fast
if the bias \eqref{EQ:bias-v} is towards the absorbing wall, (ii)
to vanish slowly if the bias is zero, and (iii) to persist indefinitely
if the walkers are driven away from the wall. Therefore $v<0$ is
the inactive phase, $v=0$ is the critical point, and $v>0$ is the
active phase.

We start our analysis by identifying the order parameter. Since the
clean critical theory of the SEI model is in the dynamical percolation
universality class, the order parameter is the percolation probability
${\cal P}$ which is linearly proportional to the survival probability,
i.e., the probability that the system has never reached $n_{IS}=0$.
In the fBMAW framework, the survival probability is the persistence
probability which, for zero bias velocity and large number of steps
$n\gg1$, is known to decay as~\citep{Krug_PersExp1997,Zoia_AsympSelfAfine2009,Wiese_PertTheoryAbsFBM2011}
\begin{equation}
{\cal P}\sim t^{-\delta},\label{EQ:Pcrit-corr}
\end{equation}
with $\delta=\gamma/2$ and $t=\Delta t_{1}+\dots+\Delta t_{n}\approx n\left[\Delta t_{n}\right]$.

We now turn our attention to $N_{IS}$. Since $\left[N_{IS}\right]\sim\left[e^{x}\right]$,
the leading behavior of $N_{IS}$ is given by the large-$x$ behavior
of the fBMAW. According to Ref.\ \citep{Wiese_FBM_Absorbing}, the
probability density of the fBMAW after $n\gg1$ steps decays as 
\begin{equation}
P_{\text{tail},n}\left(x\right)\sim\exp\left[-x^{2}/(2\sigma^{2}n^{2-\gamma})\right]\label{eq:Ptail-crit}
\end{equation}
 for large $x$ and up to a subleading power-law correction. Thus,
\begin{equation}
[N_{IS}]\sim\int_{0}^{\infty}e^{x}\exp\left[-x^{2}/\left(2\sigma^{2}n^{2-\gamma}\right)\right]\text{d}x\sim e^{At^{\theta}},\label{EQ:NIS-crit}
\end{equation}
with $\theta=2-\gamma$ and $A$ being a constant, i.e., $\left[N_{IS}\right]$
grows exponentially at the critical point. Notice that $\left[N_{IS}\right]$
grows faster than $e^{vt}$ if $\gamma<1$ which is physically impossible
since $\left[N_{IS}\right]$ can not grow faster than the growth in
the clean active phase \eqref{EQ:NIS-clean}. This accelerated growth
happens due to $\left[N_{IS}\right]$ been dominated by the arbitrarily
high noise values generated by correlated Gaussian distribution. In
a real situation the noise is bounded $\left|\xi_{n}\right|<\xi_{\text{max}}$
implying that $N_{IS}$ can be at best proportional to $e^{\xi_{\text{max}}n}$.
Setting $\xi_{\text{max}}n$ as the upper limit of the integral \eqref{EQ:NIS-crit},
we find that $\theta=\min\{1,2-\gamma\}.$

We now show that the arithmetic $\left[N_{IS}\right]\sim\left[e^{x}\right]$
and the geometric $N_{IS}^{\text{typ}}\sim e^{[x]}$ averages are
quite different. Starting at $t=0$ from $x_{0}=\ln(1+N_{IS}^{0})$
the walkers perform an unbounded fBM until they reach the absorbing
wall. The large-$x$ behavior of the surviving walkers are thus dictated
by \eqref{eq:Ptail-crit} and hence $\left[x\right]_{\text{surv}}\sim\int_{0}^{\infty}\text{d}xxP_{\text{tail},n}\left(x\right)\sim t^{1-\gamma/2}$.
As expected, it scales as the width of $P_{\text{tail},n}$. Finally,
taking into consideration the surviving probability, we find that
\begin{equation}
\ln\left(1+N_{IS}^{\text{typ}}\right)\sim\left[x\right]\sim t^{1-\gamma/2}{\cal P}\sim t^{\alpha},\label{EQ:xcrit-corr}
\end{equation}
with $\alpha=1-\gamma$. Interestingly, notice that $x$ grows if
the correlations are positive $\gamma<1$, vanishes if the correlation
are negative $\gamma>1$, and remains constant in the uncorrelated
case $\gamma=1$. As anticipated, the arithmetic and geometric averages
behave quite different at criticality as can be quantified via $\ln[N_{IS}]/\ln N_{IS}^{\text{typ}}\sim t$.
Such strong difference is a hallmark of the infinite-noise criticality.

\subsection{Off-critical properties}

After analyzing some critical properties of the system where exponents
$\delta$, $\theta$ and $\alpha$ were founded, we now turn attention
to the off-critical behavior. We start with the correlation time $\xi_{t}$.
Without resorting to the time correlation function, $\xi_{t}$ can
be evaluated as the time at which the average position of the surviving
walkers $\left[x\right]_{\text{surv}}$ crosses over from the critical
to the off-critical average position. As we have seen, at criticality
$\left[x\right]_{\text{surv}}\sim t^{(2-\gamma)/2}$. In the active
phase, the walkers are subject to a bias velocity $v>0$ and therefore
the average position grows ballistically $\left[x\right]_{\text{surv}}\propto vt$
for $t\gg1$. The crossover (correlation) time between the critical
and off-critical regimes is thus 
\begin{equation}
\xi_{t}\sim|v|^{-\nu_{\parallel}},\label{EQ:xi-offcrit-corr}
\end{equation}
with $\nu_{\parallel}=2/\gamma$, which saturates the generalized
Harris criterion \eqref{EQ:GHC}.

What about the order parameter? It is proportional to the stationary
survival probability ${\cal P}_{\text{st}}$ which can be estimated
as the critical ${\cal P}$ in \eqref{EQ:Pcrit-corr} at the crossover
(correlation) time $\xi_{t}$ in \eqref{EQ:xi-offcrit-corr}. Therefore,
\begin{equation}
{\cal P}_{\text{st}}\sim v^{\beta},\label{EQ:P-offcrit-corr}
\end{equation}
where $\beta=\delta\nu_{\parallel}=1$ is a $\gamma$-independent
critical exponent.

Can we determine the relation between the correlation length $\xi$
and the correlation time $\xi_{t}$? On the Bethe lattice, the notion
of length is not clear. Nevertheless, we can define a compact cluster
of size $\xi$ containing $N_{IS}\sim\xi^{d}$ active bonds. In the
active phase, this cluster grows ballistically in time as $\ln N_{IS}\sim v\xi_{t}$.
Therefore, together with \eqref{EQ:xi-offcrit-corr} we find that
\begin{equation}
\ln\xi\sim\xi_{t}^{\psi},\label{EQ:Activated-scaling}
\end{equation}
with ``tunneling'' exponent $\psi=1-\nu_{\parallel}^{-1}=1-\gamma/2=H$,
which equals the Hurst exponent. This activated dynamic scaling relation
between length and time scales is another hallmark of the infinite-noise
critical behavior. 

Let us remark that the exponents $\theta$, $\alpha$, $\delta$,
$\nu_{\parallel}$ and $\psi$ are $\gamma$-dependent. If one uses
another noise correlation function instead of \eqref{EQ:G-FBM} such
as $G\propto(1+\left|t\right|)^{-\gamma}$, then our results hold
for $\gamma\leq1$ (positive correlations), while for $\gamma>1$
these exponents remain the same as for the uncorrelated case $\gamma=1$.
We end this section by comparing the critical exponents here reported
to those of the contact process and its generalizations. The exponents
$\delta$, $\nu_{\parallel}$, $\beta$, and $\psi$ {[}respectively
in Eqs.~\eqref{EQ:Pcrit-corr}, \eqref{EQ:xi-offcrit-corr}, \eqref{EQ:P-offcrit-corr}
and \eqref{EQ:Activated-scaling}{]}, were also computed in those
models~\citep{VojtaHoyos15,Fiore_SecondOrderTempDis2018,AHOW_ExtinctionTransition}
and are numerically equal. Thus, the critical behavior of SEI model
here studied is in the same infinite-noise universality class of the
contact process in $d=\infty$.

\subsection{Absence of the temporal Griffiths phase}

In the contact process, the active temporal Griffiths phase~\citep{TemporalGP_Vaszquez}
is caused by rare sequences of large time intervals in which the system
is temporarily in the inactive phase while the system is actually
in the active one. As a consequence, these rare time intervals causes
strong density fluctuations near criticality. Can we observe an analogous
effect in the SEI model?

Let us analyze the probability for a walker to get absorbed in the
active phase $v>0$. In this case, the distribution of the walker
position is a Gaussian centered at $\left\langle x\right\rangle \sim vt$
and width $\sqrt{\left\langle x^{2}\right\rangle -\left\langle x\right\rangle ^{2}}\sim t^{\psi}$,
with $\psi=1-\gamma/2$. Since $\psi<1$, the probability for a particle
to get absorbed goes to zero in the long-time limit. As a consequence,
the relative density fluctuations will vanish $\sim t^{\psi-1}$ with
a prefactor that is inversely proportional to $x_{0}$. In fact, we
could not observe the active temporal Griffiths phase in our simulations
for $x_{0}\gtrsim5$. A similar argument can be applied to the inactive
temporal Griffiths phase.

\section{Monte-Carlo simulations\label{SEC:MC}}

In this section we report our Monte Carlo simulations on the SEI model
with correlated temporal disorder and compare the numerical results
with the analytical theory.

\subsection{The algorithm, the random variables and the averages}

In our simulations we start with a single cluster of $n_{I}^{0}$
infected individuals on a lattice full of susceptible ones ($n_{E}^{0}=0$,
and thus $n_{IS}^{0}=\left(z-2\right)n_{I}^{0}+2$). As explained
in Sec.~\ref{SEC:SEI-def}, activity occurs at the border of this
cluster and the lattice is irrelevant for the simulations. Therefore,
only the number of $n_{IS}$ pairs is necessary for the simulation.
Let $r_{1}$ and $r_{2}$ be uniformly distributed random numbers
in the interval $]0,1[$, and $p_{n}=\lambda_{n}/(\lambda_{n}+\mu_{n})$.
Our algorithm then reads: 
\begin{enumerate}
\item increase time $t$ by $1/(n_{IS}r_{1})$. 
\item generate new values of $\mu_{n}$ and $\lambda_{n}$ if the integer
part of the discrete time $n=t/\Delta t$ has increased by one unity.\footnote{Without loss of generality, we assume $\Delta t_{n}=\Delta t$.}
\item if $r_{2}<p_{n}$, an \emph{S} site becomes $I$. This means that
we need to increase the number of $IS$ pairs by $z-2$. On the other
hand, if $r_{2}\geq p_{n}$, the $S$ site becomes $E$ which decreases
the number of \emph{IS} pairs by one. 
\item return to (i) until $n_{IS}=0$. 
\end{enumerate}
However, despite this algorithm being of easy implementation and allow
us to simulate system sizes as large as the computer precision, it
is very slow for large $n_{IS}$ since the time increasing is, on
average, only $\sim1/n_{IS}$ at each iteration. In other words, the
simulation time increases exponentially with $n_{IS}$. 

In order to optimize the SEI simulations, we will use a combination
of the above mentioned algorithm and the analytical result \eqref{EQ:NIS-dirty}.
First, we introduce a cutoff $N_{IS}^{\text{cutoff}}=10^{3}$. When
$n_{IS}<N_{IS}^{\text{cutoff}}$, $n_{IS}$ is small and we can use
the SEI Monte Carlo algorithm above described. If $n_{IS}>N_{IS}^{\text{cutoff}}$,
$n_{IS}$ is very large and we expect the stochastic noise to average
out. Therefore, $n_{IS}$ should be well described by its stochastic
average $N_{IS}$.\footnote{For this reason, $n_{IS}\geq0$ is treated as a continuous variable
and the simulation survives while $n_{IS}>0$.}

Having described the SEI simulation algorithm, we now explain the
generation of power-law correlated noise. Following Refs.\ \onlinecite{AHOW_ExtinctionTransition,AHOW_FBM,WadaFBM_2019},
we employ the Fourier filtering method~\citep{Makse_1996} to generate
Gaussian correlated noise following the correlation function \eqref{EQ:G-FBM}.
We first generate a set $\nu_{n}$ of zero average Gaussian uncorrelated
numbers of variance $1$. The Gaussian power-law correlated random
numbers $\zeta_{n}$ are given by the inverse Fourier transform of
\begin{equation}
\tilde{\zeta}_{\omega}=[\tilde{G}_{\text{fBM}}(\omega)]^{1/2}\tilde{\nu}_{\omega},
\end{equation}
where $\tilde{\nu}_{\omega}$ and $\tilde{G}_{\text{fBM}}$ are the
Fourier transform of $\nu_{n}$ and of the correlation function \eqref{EQ:G-FBM},
respectively. A noise of variance $\sigma^{2}$ can be obtained by
multiplying $\zeta{}_{n}$ by $\sigma$.

In the following simulations we fix $\lambda_{n}=\lambda$ and $\mu_{n}=\mu+\Delta\mu_{n}$
is the source of temporal disorder, i.e., $\Delta\mu_{n}$ is a zero-mean
Gaussian noise with variance $\sigma^{2}$. To control the distance
to the critical point $v$ \eqref{EQ:bias-v}, we choose $\mu=128$
(sufficiently large ensuring that $\mu_{n}>0$ in our simulations)
and vary only $\lambda$. All simulations use the coordination number
$z=4$ (so the critical point $v=0$ happens for $\lambda=64$) and
run up to $t=10^{8}$. Furthermore, for convenience, time will always
be renormalized by a factor of $1/(\mu+\lambda)$. Therefore, for
a small time window $\delta t$, $n_{IS}\delta t$ processes take
place on average.

\begin{figure}
\centering{}\includegraphics{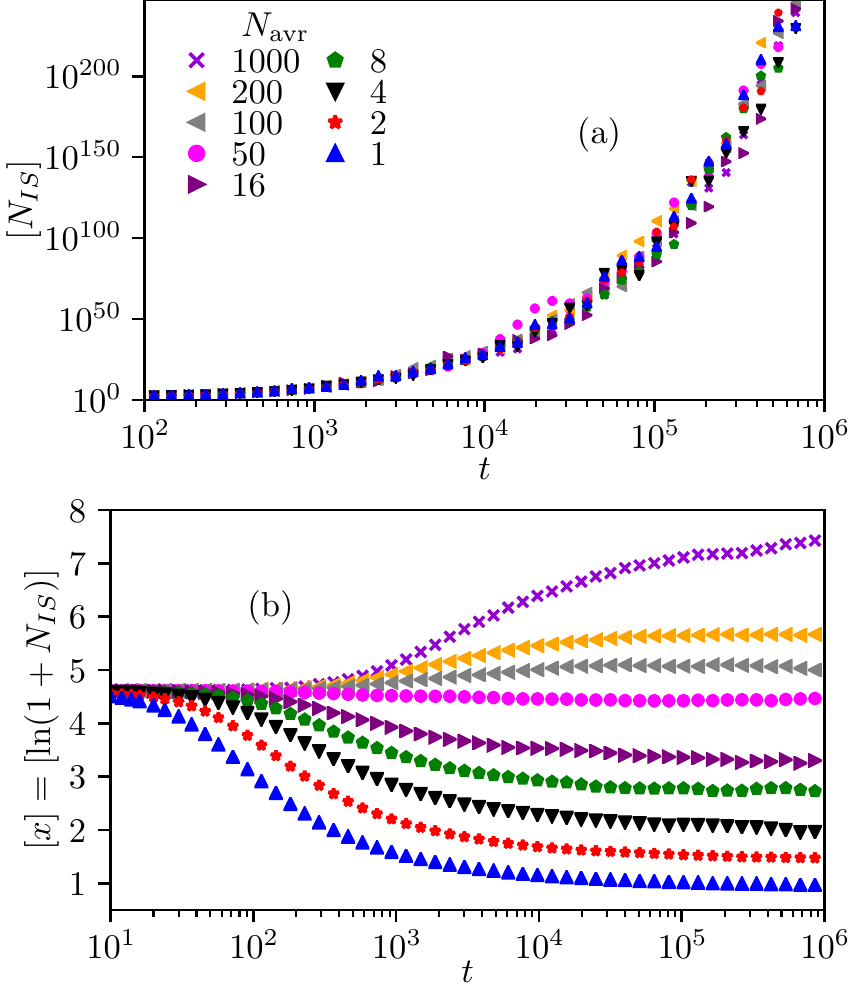} \caption{Simulations of the SEI model with uncorrelated temporal disorder ($\gamma=1$)
at the critical point $v=(z-2)\lambda-\mu=0$. Here $\lambda=64$,
$\mu=128$, $\sigma=8$, $\Delta t=16,$ the initial cluster has size
$N_{IS}^{0}=10^{2}$, and the data is averaged over ${\cal N}=10^{5}$
temporal disorder configurations and over $N_{\text{avr}}$ realizations
of the stochastic noise. In panel (a) the uncertainties are of the
order of the dispersion of the data and in panel (b) of the order
of the symbol size.\label{FIG:FiniteX}}
\end{figure}

Finally, let us discuss about the combined stochastic and external
noise averages $\left[\left\langle {\cal A}\right\rangle \right]$
of a certain observable ${\cal A}$. The estimates of the stochastic
$\left\langle {\cal A}\right\rangle $ and the external noise $\left[{\cal A}\right]$
averages are taken over $N_{\text{avr}}$ stochastic noise and ${\cal N}$
temporal disorder realizations, respectively. Interestingly, we find
in our simulations that near criticality $N_{\text{avr}}$ can be
as low as $1$ and ${\cal N}\gg1$. This is the first confirmation
of our theory. The infinite noise critical point is characterized
by long incursions of the system in either of the phases around the
transition point yielding to ever larger fluctuations. In other words,
effectively the system is never under the clean critical point influence,
but fluctuating around and far from it. This is seen in Fig.~\ref{FIG:FiniteX}.
In panel \hyperref[FIG:FiniteX]{(a)}, $\left[\left\langle n_{IS}\right\rangle \right]$
is shown as a function of time for the case of uncorrelated disorder
$\gamma=1$ and at criticality $\mu=\mu_{c}$. We have averaged over
${\cal N}=10^{5}$ temporal disorder configurations and over different
values of stochastic noise $N_{\text{avr}}$ ranging from as low as
$N_{\text{avr}}=1$ up to $N_{\text{avr}}=10^{3}$. Clearly, $\left[N_{IS}\right]$
is barely sensitive to $N_{\text{avr}}$ within the temporal disorder
statistical error. Likewise, we plot $\left[x\right]=\left[\ln\left(1+N_{IS}\right)\right]$
in panel \hyperref[FIG:FiniteX]{(b)} as a function of time. The effects
of different values of $N_{\text{avr}}$ is only quantitative higher
since $\left[x\right]\rightarrow\text{const}$ as $t\rightarrow\infty$.
The reason for this stronger effect is because the typical average
is not dominated by tail of the probability distribution as is $\left[N_{IS}\right]$.
Since neither $\left[\left\langle n_{IS}\right\rangle \right]$ nor
$\left[\ln\left\langle 1+n_{IS}\right\rangle \right]$. are qualitatively
affected by $N_{\text{avr}}$ we will take $N_{\text{avr}}=1$ from
now on.

\subsection{Critical properties}

Having discussed the details of our simulations, we now report on
our numerical results.

\begin{figure}
\centering{}\includegraphics{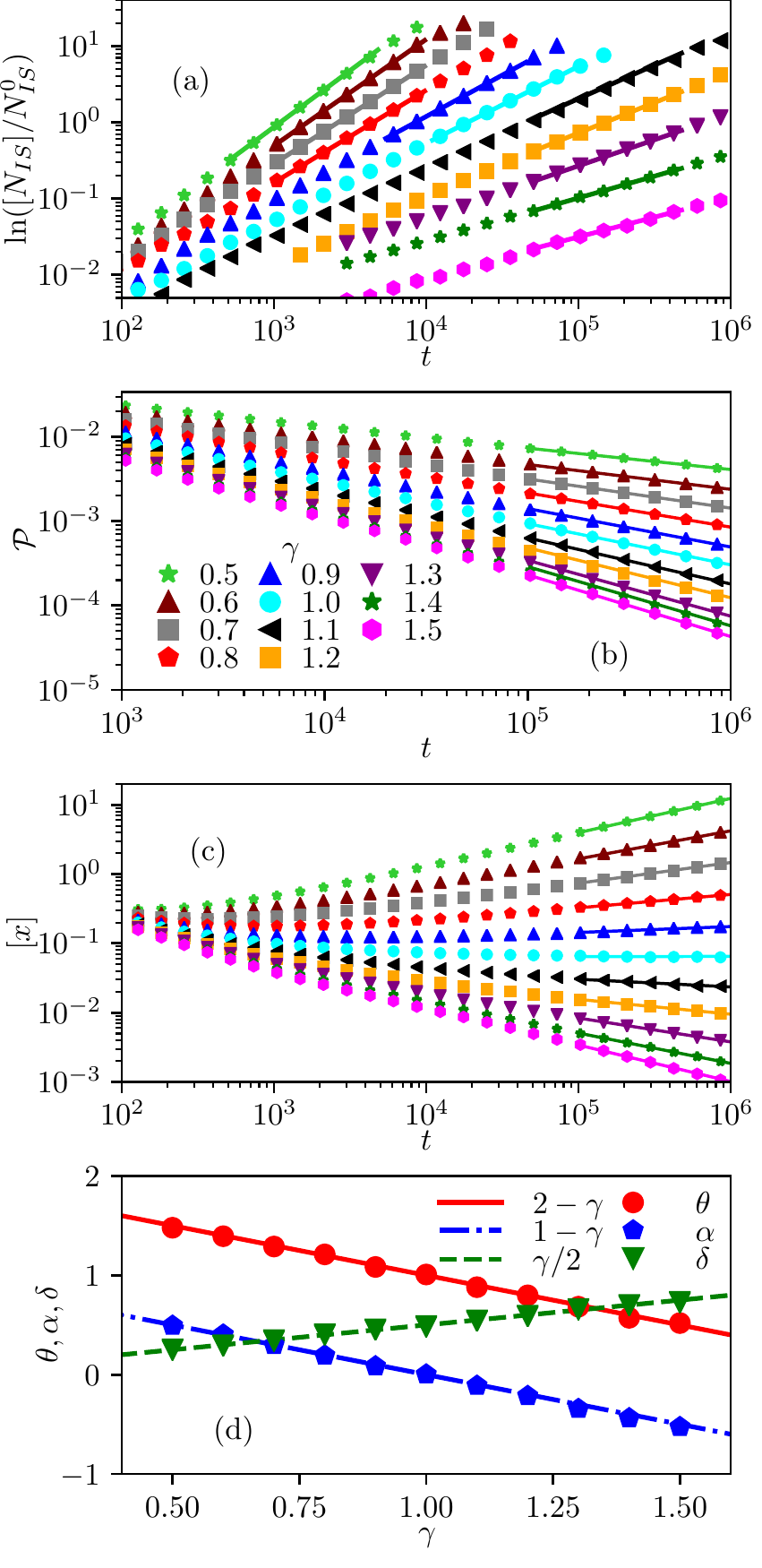}\caption{Simulations of the SEI model with power-law correlated temporal disorder
at the critical point. In all panels the uncertainties are of the
order of the symbol size, and the solid lines are power-law fits.
(a) $\ln([N_{IS}]/N_{IS}^{0}$) vs.\ $t$. for many values of the
correlation exponent $\gamma$ {[}see Eq.~\eqref{EQ:G-FBM}{]}. These
simulations used the parameters $\sigma=1$, $\Delta t=4$ and initial
condition $N_{IS}^{0}=4$. (b,c) the survival probability ${\cal P}$
and the average position $[x]=\ln(1+N_{IS})$ as a function of time
for $\sigma=8$, $\Delta t=16$ and initial condition $N_{IS}^{0}=4$.
(d) The critical exponents extracted by fitting power laws to the
data according to the equations \eqref{EQ:NIS-crit}, \eqref{EQ:Pcrit-corr}
and \eqref{EQ:xcrit-corr}, for $N_{IS}$, ${\cal P}$ and $x$, respectively.
All simulations averages over $10^{7}$ disorder configurations.\label{FIG:Crit}}
\end{figure}

Figure \hyperref[FIG:Crit]{\ref{FIG:Crit}(a)} shows our simulations
for $\ln([N_{IS}]/N_{IS}^{0})$ at the critical point for many values
of $\gamma$. In this figure we can verify that $\ln([N_{IS}]/N_{IS}^{0})$
follows power laws for many decades in the $y$ axis in agreement
with $\ln[N_{IS}]\sim t^{\theta}$ Eq.~\eqref{EQ:NIS-crit}. Furthermore,
power-law fits to the data (solid lines) yield the exponents plotted
in Fig.~\hyperref[FIG:Crit]{\ref{FIG:Crit}(d)} (red circles), which
are compatible with our prediction $\theta=2-\gamma$. Notice, however,
that our data shows deviation from the power-law behavior for large
$\ln([N_{IS}]/N_{IS}^{0})$. Since the leading behavior of $\ln([N_{IS}]/N_{IS}^{0})$
is given by the large-$x$ tail of the probability density (rare fluctuations),
we argue that this deviation is due the lack of a larger number of
temporal disorder configurations when performing the average. We report
that increasing $N_{\text{avr}}$ does not change this feature.

Now we turn our attention to the survival probability ${\cal P}$
(estimated as the fraction of survival runs at time $t$) and the
average position $[x]$, which can be seen in Figs.~\hyperref[FIG:Crit]{\ref{FIG:Crit}(b)}
and \hyperref[FIG:Crit]{(c)} as a function of time. In comparison
to $\ln[N_{IS}]$, the plots of both ${\cal P}$ and $[x]$ are smoother
as expected since the leading contribution does not come from the
rare regions. The solid lines are simple power-law fits to our predictions
${\cal P}\sim t^{-\delta}$ \eqref{EQ:Pcrit-corr} and $[x]\sim t^{\alpha}$
\eqref{EQ:xcrit-corr}. The corresponding exponents $\delta$ (green
triangles) and $\alpha$ (blue pentagons) are shown in Fig.~\hyperref[FIG:Crit]{\ref{FIG:Crit}(d)}
alongside the analytical expectation (solid lines) where agreement
is remarkable.

\subsection{Off-critical properties}

We now report our simulations in the active phase $v=(z-2)\lambda-\mu>0$
{[}see Eq.~\eqref{EQ:bias-v}{]} and evaluate the exponents $\beta$
and $\nu_{\parallel}$.

\begin{figure}
\centering{}\includegraphics{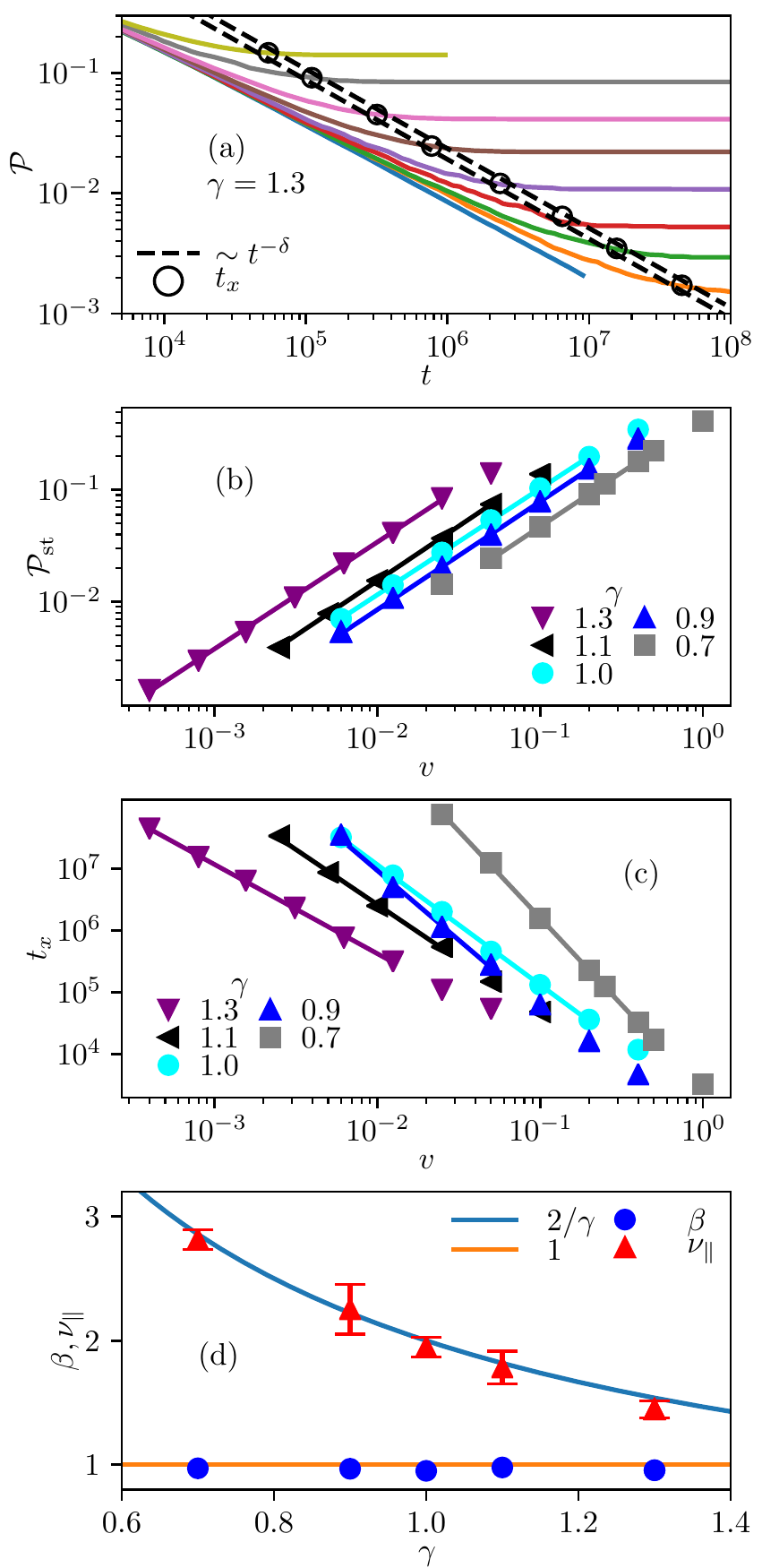}\caption{Off-critical properties of the SEI model. Uncertainties, if not shown
and except for $\beta$, are of the order of the symbol size. The
temporal disorder strength is $\sigma=8$ and the time interval is
$\Delta t=16$. (a) The survival probability (solid lines) in the
active phase and at the critical point for $\gamma=1.3$. The bias
velocity $v$ from bottom to top are $0$, $0.0004$, $0.0008$, $0.0015625$,
$0.003125$, $0.00625$, $0.0125$, $0.025$ and $0.05$. The black
dashed curves represents the conficence interval of $P_{\text{{crit}}}$
vertically shifted which intercept the off-critical curves at the
crossover time $t_{x}$ (black circles). (b) The stationary survival
probability $P_{\text{st}}$ and (c) the crossover time $t_{x}$ vs.
the distance to the critical point $v$ for many values of $\gamma$.
The solid lines are power-law fits. (d) The critical exponents $\beta$
and $\nu_{\parallel}$ for many values of $\gamma$. The solid lines
are the predictions to $\beta$ and $\nu_{\parallel}$ Eqs.~\eqref{EQ:P-offcrit-corr}
and \eqref{EQ:xi-offcrit-corr}. The uncertainties of $\beta$ are
of order $10^{-3}.$ The data are averaged over $10^{7}$ disorder
configurations. \label{FIG:OffCrit}}
\end{figure}

Figure \hyperref[FIG:OffCrit]{\ref{FIG:OffCrit}(a)} shows our simulations
of the survival probability ${\cal P}$ at the critical point and
in the active phase. At the critical point, the survival probability
decays as power law according to ${\cal P}\sim t^{-\gamma/2}$ \eqref{EQ:Pcrit-corr}.
In the active phase, $v>0$, ${\cal P}$ reaches a stationary state
${\cal P}_{\text{st}}$ after a transient time, meaning that some
(percolating) clusters generated by the SEI dynamics survive indefinitely.

The stationary survival probability ${\cal P}_{\text{st}}$ can be
estimated by fitting a constant at the asymptotic behavior of $\mathcal{P}$
for the curves far enough from the critical point, and are plotted
in Fig.~\hyperref[FIG:OffCrit]{\ref{FIG:OffCrit}(b)} as a function
of $v$. By fitting power laws to the data we are able to estimate
the exponent $\beta$ which is plotted as blue circles in Fig.~\hyperref[FIG:OffCrit]{\ref{FIG:OffCrit}(d)}.
Our values of $\beta$ show deviations, roughly around $0.02$, from
the prediction in Eq.~\eqref{EQ:P-offcrit-corr} $\beta=1$ that
is not within the uncertainty bar (of order $10^{-3})$. We believe
that this deviation occurs due to the asymptotic regime is not completely
reached and ever larger times are needed.

The correlation time $\xi_{t}$ for a given bias $v$ can be estimated
roughly as the crossover time $t_{x}$ where the off-critical curve
crosses over to a curve proportional to the critical curve $P_{\text{crit}}$.
The black dashed lines in Fig.~\hyperref[FIG:OffCrit]{\ref{FIG:OffCrit}(a)}
shows the confidence interval of $P_{\text{{crit}}}$ (multiplied
by a constant) and the empty circles the estimated values of $t_{x}$.

In Fig.~\hyperref[FIG:OffCrit]{\ref{FIG:OffCrit}(c)} we plot $t_{x}$.
Deviations from a simple power law for larger values of $v$ are due
to corrections to scaling. The best power-law fits yields our estimates
of $\nu_{\parallel}$ which are plotted in Fig.~\hyperref[FIG:OffCrit]{\ref{FIG:OffCrit}(d)}
as red triangles and agrees with our prediction $\nu_{\parallel}=2/\gamma$
Eq.~\eqref{EQ:xi-offcrit-corr}.

\section{The compact directed percolation and the infinite-noise universality
classes\label{sec:CDP}}

In this section we uncover a relation between the $d=\infty$ infinite-noise
and the compact directed percolation (CDP) universality classes. 

\begin{figure}
\begin{tikzpicture}[scale=0.5]
	\foreach \x in {0, 2, 4, 6}{
		\draw (\x,0) circle (0.25);
		\draw (\x,2) circle (0.25);
		\draw (\x+1,1) circle (0.25);
		\draw[dotted] (\x+.2,.2)--(\x+0.8, 0.8);
		\draw[dotted] (\x+.2,1.8)--(\x+0.8, 1.2);
		\draw[dotted] (\x+1.2,0.8)--(\x+1.8, 0.2);
		\draw[dotted] (\x+1.2,1.2)--(\x+1.8, 1.8);
	}
	\draw (8,0) circle (0.25);
	\draw (8,2) circle (0.25);

	\node[anchor=south] at (2,2.25) {\scriptsize $i-1$};
	\node[anchor=south] at (3,1.25) {\scriptsize $i$};
	\node[anchor=south] at (4,2.25) {\scriptsize $i+1$};
	\draw[->, anchor=east,thick] (-0.75, 3) -- (9, 3);

	\node[anchor=east] at (-1, 2) {\scriptsize $t  $};
	\node[anchor=east] at (-1, 1) {\scriptsize $t+1$};
	\node[anchor=east] at (-1, 0) {\scriptsize $t+2$};
	\draw[->, anchor=east,thick] (-0.75, 3) -- (-0.75, -0.5);

\end{tikzpicture}\caption{Tilted lattice of the Domany-Kinzel cellular automaton. The horizontal
axis represents the index $i$ of the lattice sites and the vertical
is the time $t$. Notice that all site indices are even (odd) for
even (odd) time slices $t$.\label{fig:Tilded-lattice}}
\end{figure}
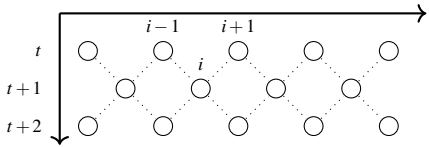

As a brief introduction to the CDP universality class~\citep{HenkelHinrichsenLuebeck_book08},
consider the Domany-Kinzel cellular automaton \citep{CDP_DomanyKinzel1984}
which is defined on a tilted lattice (see Fig. \ref{fig:Tilded-lattice})
where a site at time $t$ can be active or inactive. The probability
that the $i$th site will be occupied at time $t+1$ is $p_{1}$ if
only one of the neighbors ($i-1$ or $i+1$) is active at time $t$,
or $p_{2}$ if both neighbors are active. Otherwise the $i$th site
remains inactive. For $p_{2}<1$ there is a phase transition in the
directed percolation universality class.

For $p_{2}=1$, on the other hand, the transition is in the CDP universality
class due to the additional particle-hole symmetry. In this case,
inactive (active) sites inside a cluster of active (inactive) ones
are not allowed and, therefore, the clusters are compact and the dynamics
occurs only at the interface between active and inactive particles.
For $p_{1}>p_{1,c}=1/2$ ($p_{1}<p_{1,c}$), the clusters of active
(inactive) sites increase and the system reaches an absorbing state
of active (inactive) sites. At the transition point $p_{1,c}$ the
interface performs an unbiased random walk.

By starting with a single cluster of active particles, this special
case $p_{2}=1$ of the Domany-Kinzel model can be easily solved \citep{CDP_DomanyKinzel1984,CDP_Essam1989}
since the number of active particles $x$ (which is now the difference
between the position of the right and left interfaces) also performs
a random walk with an absorbing wall at $x=0$. Therefore, this special
limit of the one-dimensional Domany-Kinzel model shares many similarities
to the SEI model with uncorrelated temporal disorder on the Bethe
lattice. The stochastic noise in the former plays the role of uncorrelated
($\gamma=1$) temporal disorder in the latter. As we have shown, the
details of the absorbing wall is irrelevant for the critical properties
and thus, we conclude that both models are governed by the same critical
point.

In fact, the critical exponents of the CDP universality class in $d=1$
confirm this statement (see Table \ref{tab:Critical-exponents}).
The survival probability ${\cal P}$ and the average position $[x]$
are equivalent in both models, and thus, there is a perfect correspondence
of the exponents $\delta$, $\nu_{\parallel}$, $\beta$ and $\alpha$.
In the cellular automaton, the critical dynamical exponent $z$ is
defined from the relation between length and time scales $\xi^{\text{CDP}}\sim\left(\xi_{t}^{\text{CDP}}\right)^{z},$
with $z=1/2$. Noting that $\xi^{\text{CDP}}$ is the equivalent of
$\ln N_{IS}\sim\ln\xi$ in Eq.~\eqref{EQ:Activated-scaling}, then
the exponent $z$ of CDP is equivalent to $\psi$ in the SEI model.

Having shown the similar critical behavior of these models, we now
recall that the usual hyperscaling relation is violated by the CDP
critical exponents because the system exhibits more than one absorbing
state~\citep{DK_Hyperscaling_Dickman1995}. Instead, a more general
relation is obeyed: $dz=\alpha+\left(\beta+\beta^{\prime}\right)/\nu_{\parallel}$.
Here, $\beta^{\prime}$ is defined via $\rho\sim\left(p_{1}-p_{1,c}\right)^{\beta^{\prime}}$,
where $\rho$ is the density of active sites in the stationary state.
For $p_{1}<p_{1,c}$, it is clear that $\rho=0$ since the inactive
state is the only stable absorbing state. Particle-hole symmetry then
demands that $\rho=1$ in the active phase and thus, $\rho$ is discontinuous
at the transition. For this reason and taking $\rho$ as the order
parameter, the CDP transition is discontinuous and $\beta^{\prime}=0$.
Since the density $\rho$ is proportional to the size of the active
cluster $x$, then the equivalent quantity in the SEI model is simply
$\left[x\right]$. As its stationary value is discontinuous at the
transition for any $\gamma$, we then conclude that $\beta^{\prime}=0$
in our model. Finally, we arrive at the corresponding hyperscaling
relation 
\begin{equation}
\psi=\alpha+\frac{\beta+\beta^{\prime}}{\nu_{\parallel}},\label{eq:Hyperscaling}
\end{equation}
 which is satisfied for all $\gamma$ (see Table \eqref{tab:Critical-exponents}).

We finally end this section by pointing out that our results straightforwardly
apply to a generalization of the $d=1$ CDP universality class. Here,
if one replaces the usual white stochastic noise by a fractional Gaussian
one (or by a generic long-range correlated noise), the critical exponents
change accordingly as in Table \eqref{tab:Critical-exponents}.

\begin{table}
\begin{tabular}{c|C{7mm}|C{7mm}|C{7mm}|C{7mm}|C{7mm}|C{10mm}}
               & $\beta'$ & $\nu_\parallel$ & $\beta$ & $\alpha$ & $\delta$   & $z$ (or $\psi$) \\ \hline
CDP       & $0$ & $2$  & $1$ &  $0$   & $1/2$     & $1/2$    \\
this work & $0$  & $2/\gamma$ & $1$ & $1-\gamma$ & $\gamma/2$ & $1-\gamma/2$ \\ \hline
\end{tabular}

\caption{The critical exponents of the $d=1$ compact directed percolation
(CDP)~\citep{CDP_DomanyKinzel1984,CDP_Essam1989,DK_Hyperscaling_Dickman1995}
and of the $d=\infty$ (fractional Gaussian power-law correlated)
infinite-noise universality classes. \label{tab:Critical-exponents}}
\end{table}

\section{Conclusion \label{SEC:Conclusion}}

We have studied the phase transition of the SEI model on the Bethe
lattice and in the presence of time-varying power-law correlated noise
(temporal disorder) $G(t)\sim\left(1-\gamma\right)t^{-\gamma}$ {[}see
Eq.~\eqref{EQ:G-FBM}{]}. By starting with a cluster of infected
sites on the lattice full of susceptible individuals we could write
exact equations for the time evolution of the active bonds $N_{IS}$
and map the problem onto a fractional Brownian motion $x\sim\ln N_{IS}$
with an absorbing wall (fBMAW) at $x=0$ when the number of active
bonds $N_{IS}$ is sufficiently large.

The mapping onto a fBMAW allowed us to completely characterize the
critical behavior analytically. At the criticality, the distribution
of $x$ broadens without limit with increasing time implying that
the typical and average values are quite different: $\ln[N_{IS}]/\ln N_{IS}^{\text{typ}}\sim t\rightarrow\infty$,
and thus, the system critical behavior is of the infinite-noise character.
In addition, we have determined that the corresponding critical exponents
are equal to those governing the critical properties of the contact
process with correlated temporal disorder in $d=\infty$~\citep{AHOW_ExtinctionTransition}.
Therefore, the mean-field criticality of the SEI model with temporal
disorder is of infinite-noise type and in the same universality class
of the contact process. However, the off-critical behavior of these
models are quite different with absence of a Griffiths phase in the
former model. The technical reason is that the boundary condition
plays an important role off criticality. While the confining reflecting
wall allows for large fluctuations of the walker in the contact process,
the unconfined absorbing wall prevents large fluctuations in the SEI
model.

It is interesting to draw a parallel between the infinite-noise and
the infinite-randomness universality classes in seemingly unrelated
models. The infinite-randomness critical behavior is observed, for
example, in the (quantum) random-transverse field Ising chains with
spatially power-law-correlated quenched disorder~\citep{Rieger_PRB1998,Rieger_1999}.
In this case, it was shown that (i) the critical surface magnetization
scales with $[m_{s}]\sim L^{-\gamma/2}$, that (ii) the off-critical
(spatial) correlation length behaves according to $\xi\sim v^{-2/\gamma}$,
and that (iii) the logarithm of energy gap scales with system size
$L\sim(\ln\Delta E)^{2/(2-\gamma)}$. These are exactly our Eqs.~\eqref{EQ:Pcrit-corr},
\eqref{EQ:xi-offcrit-corr} and \eqref{EQ:Activated-scaling} with
the roles of time and space exchanged. (Similar analogy applies for
the contact process with spatial quenched disorder~\citep{Ibrahim_longrangeDisCP}.)

Finally, we have uncovered a relation between the SEI model with temporal
disorder on the Bethe lattice and the $d=1$ Domany-Kinzel cellular
automaton model in the limit of compact clusters. It allowed us to
show the equivalence between the compact directed percolation and
the infinite-noise universality classes. 

\section{Acknowledgment}

This work was supported by the São Paulo Research Foundation (FAPESP)
under Grants No. 2015/23849-7, No. 2016/10826-1 and No. 2018/25441-3,
and CNPQ under Grant No. 312352/2018-2.

\bibliographystyle{apsrev4-1}
\bibliography{Paper.bib}
\end{document}